\documentclass[10pt, conference]{IEEEtran}
\usepackage{algorithmicx}
\usepackage[ruled,vlined,linesnumbered]{algorithm2e}
\usepackage{hhline}
\usepackage{amsmath,mathtools}
\usepackage{amsfonts,amssymb}
\usepackage{mathrsfs}
\usepackage{gensymb} 
\usepackage{caption}
\usepackage{multirow}
\usepackage{graphicx} 
\usepackage{multirow}
\usepackage{enumitem,color}
\usepackage{algpseudocode}

\usepackage{subfigure}

\usepackage{titlesec}
\titlespacing*{\section}{1pt}{0.5ex}{0.5ex}
\titlespacing*{\subsection}{1pt}{0.5ex}{0.5ex}
\titlespacing*{\subsubsection}{1pt}{0.5ex}{0.5ex}

\begin{document}
%
\title{\huge \emph{InSlicing}:  Interpretable Learning-Assisted Network Slice Configuration in Open Radio Access Networks}

\author{\IEEEauthorblockN{Ming Zhao, Yuru Zhang, Qiang Liu \vspace{-0.2in}}\\
\IEEEauthorblockA{
University of Nebraska-Lincoln\\
qiang.liu@unl.edu}\vspace{-0.3in}
\and
\IEEEauthorblockN{Ahan Kak, Nakjung Choi \vspace{-0.2in}}\\
\IEEEauthorblockA{
Nokia Bell Labs\\
nakjung.choi@nokia-bell-labs.com}\vspace{-0.3in}
}

\maketitle

\begin{abstract}
Network slicing is a key technology enabling the flexibility and efficiency of 5G networks, offering customized services for diverse applications. However, existing methods face challenges in adapting to dynamic network environments and lack interpretability in performance models. In this paper, we propose a novel interpretable network slice configuration algorithm (\emph{InSlicing}) in open radio access networks, by integrating Kolmogorov-Arnold Networks (KANs) and hybrid optimization process. 
On the one hand, we use KANs to approximate and learn the unknown performance
function of individual slices, which converts the blackbox optimization problem.
On the other hand, we solve the converted problem with a genetic method for global search and incorporate a trust region for gradient-based local refinement.
With the extensive evaluation, we show that our proposed algorithm achieves high interpretability while reducing 25+\% operation cost than existing solutions.
\end{abstract}


\section{Introduction}
\label{sec:introduction}





%

%

%

Network slicing is one of the key technologies driving the development of 5G \cite{foukas2017network}, enabling customized network services for various mobile applications, such as autonomous driving \cite{campolo20175g}, AR/VR \cite{wijethilaka2021survey}, and IoT \cite{qiu2020edge}. 
By virtualizing physical network resources into multiple independent logical networks \cite{afolabi2018network}, multiple network slices can be concurrently supported and operated with tailored network performance and functionality, such as low-latency, high-throughput, and ultra reliability. 
With the advance of open radio access networks (e.g., O-RAN \cite{polese2023understanding}), the disaggregated network architecture, complex network dynamics, and diversified RAN Intelligent Controller (RIC) creates ever-complicating slice configuration in next-generation mobile networks.

For mobile network operators (MNOs), the key challenge is dynamic slice configuration, which aims to minimize the operational cost (e.g., resource usage) while meeting slice users’ performance requirements to assure their service level agreements (SLAs)~\cite{salvat2018overbooking}. 
With the ever-complicating mobile networks, however, it is impractical (if not impossible) to obtain the accurate representation of the complex relationship between the slice configuration and resulting slice performances~\cite{liu2022atlas, shi2021adapting}.
Hence, recent works~\cite{li2018deep, liu2020deepslicing, liu2021onslicing} focused on leveraging the superior approximation capability of AI/ML techniques, especially deep neural networks (DNNs), to learn the relationship and then optimize the slice configuration accordingly. 
For example, Atlas~\cite{liu2022atlas} used Bayesian optimization to online learn and allocate multi-dimensional resources (e.g., virtual radio bandwidth) to concurrent network slices, by combining Bayesian neural network (BNN) and Thompson sampling.
Some works~\cite{ayala2023edgebol, ayala2021bayesian} targeted the joint orchestration of both radio and computing resources, by designing a new Bayesian learning algorithm, in virtual radio access networks (vRANs).
However, existing works can only obtain the implicit representation (mostly parameterized by blackbox DNNs) of unknown correlations, which lack the explainability and interpretability of their solutions, such as \emph{why this action is taken, not others}.
As a result, although existing works could achieve promising performance, their limited interpretability remains various concerns to MNOs and eventually constrains their practical deployment in real-world networks.

In this paper, we propose a novel interpretable network slice configuration algorithm (\emph{InSlicing}) in open radio access networks.
The fundamental idea is to integrate Kolmogorov-Arnold Networks (KANs \cite{liu2024kan}) with high interpretability into the conventional non-convex optimization process. 
Distinct from traditional DNNs, KANs offer significant advantages (e.g., accuracy and interpretability) and enable intuitive visualization of how different components influence performance metrics through activation functions, providing enhanced model transparency. 
First, we formulate the slice configuration optimization problem to minimize the operation cost while satisfying the performance requirement of network slices.
Second, we design KANs to approximate and learn the unknown performance function of individual network slices.
Third, with the learned KANs, we derive the closed-form representations for individual slices, which converts the original blackbox optimization problem.
Fourth, we solve the converted problem with a population-based genetic method (GA \cite{mitchell1998introduction}) for global exploration, complemented by trust region \cite{conn2000trust} for gradient-information local refinement. 
With the extensive evaluation, we show that our proposed algorithm achieves high interpretability while reducing 25+\% operation cost than existing solutions.
We believe this work provides a new perspective of slice configuration, by considering the solution interpretability beyond the solution performance only.




\section{System Model}
We consider a generic cellular network, including the core network (CN), the radio access network (RAN) with multiple base stations and network slices. 
To assure the slice SLAs, MNOs optimize diverse configurations (e.g., resources and settings) for concurrent slices. 
Slice tenants deploy various applications within their respective slices, which impose specific performance requirements. 
From the network perspective, the performance of each slice is intrinsically related to the slice configuration decisions, denoted as \( X \). 
Each slice exhibits a unique performance function \( P(X) \). 
However, the performance function \( P(X) \) remains generally unknown, due to the complex network dynamics and unique characteristics. 
Meanwhile, slice configuration decisions incur operation costs for the MNOs.

We define \( I \) as the set of network slices, where \( i \in I \) represents the \( i \)-th slice, and \( R \) as the set of resource types, where \( r \in R \) represents the \( r \)-th resource type. Let \( x_{i,r} \) denote the amount of resource \( r \) allocated to slice \( i \). Subsequently, we can formulate the MNOs' total cost function $C(\cdot)$ as:
\begin{equation}
C(w_r , x_{i,r}) = \sum_{i \in I} \sum_{r \in R} w_r \cdot x_{i,r},
\end{equation}
where $w_r$ denotes the cost coefficient per unit of resource $r$.

The objective of network slice configuration can be formulated as a cost minimization problem, where the MNOs seek to minimize operational expenditure while ensuring that the performance requirements of slice tenants are satisfied. Specifically, the network slice configuration problem can be mathematically formulated as a constrained optimization problem as follows:
\begin{align}
    \mathbb{P}:& \quad \min_{x_{i,r}} \quad C(w_r , x_{i,r}) \\
    \text{s.t.} \quad 
    &C1: \quad  P(X_i) \ge Q_i,\quad \forall i \in I, \\
    &C2: \quad  L_{r} \le x_{i,r} \le H_r, \quad \forall i \in I, \ \forall r \in R,\\
    &C3: \quad  \sum\nolimits_{i\in I} x_{i,r} \le H_r, \quad \forall r \in R.
    \label{opt_p}
\end{align}
In problem $\mathbb{P}$, $C1$ constraints represent that the performance \( P(X_i) \) of the \( i \)-th slice, when the slice configuration action \( X_i \) is applied, should satisfy the threshold requirements \( Q_i \). Here, \( X_i = \{x_{i,1}, \ldots, x_{i,r}\} \) is the action vector allocated to slice \( i \), including all relevant resources. $C2 $ constraints ensure that the control actions \( x_{i,r} \) for each resource type \( r \) remain within the lower bound \( L_r \) and upper bound \( H_r \). $C3$ constraints impose a global restriction on resource utilization. For resource type \( r \), the total amount of resources allocated across all slices must not exceed the maximum resource capacity \( H_r \).  

\textbf{Challenge.} The primary challenge in solving the aforementioned problem $\mathbb{P}$ lies in the performance function \( P(X) \). 
On the one hand, the performance function \( P(X) \) is unknown in advance, i.e., no closed-form representation, which results in blackbox problem.
Although several frameworks (e.g., Bayesian optimization and multi-armed bandit) could solve the problem, they lack interpretability and explainability. 
On the other hand, the complex performance functions are nearly non-convex, the problem is still non-convex even if we obtain these closed-form representations.


\section{Solution}





\begin{figure}[!t]
	\centering
	\includegraphics[width=3.4in]{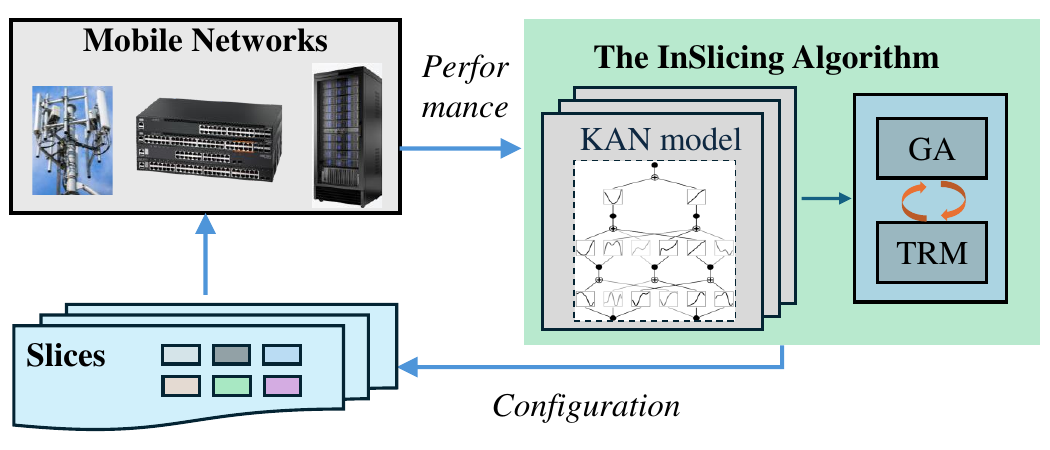}
	\vspace{-0.07in} \caption{\small The overview of the \emph{InSlicing} algorithm.}
	\label{fig:solution_fig}
\end{figure}

In this section, we present the proposed solution to effectively solve the above problem (see Fig. \ref{fig:solution_fig}), which optimizes resource usage through a combination of approximation techniques and heuristic search methodologies. 
On the one hand, we leverage KANs to perform online approximation learning of slice-specific performance functions. This approach focuses on the relationship between performance metrics and slice configuration actions for each slice, thereby transforming the previously intractable performance constraints into knowable and solvable components of the optimization problem. 
On the other hand, we solve the converted problem by developing a heuristic search method, based on genetic methods augmented with a local optimization method. 
This hybrid approach minimizes the MNOs' cost by conducting a population-based heuristic exploration of the entire resource space for each slice and evaluating slice performance. 
The proposed method is suitable for tackling complex, discontinuous, and non-convex optimization problems, which are common in slice configuration tasks with intricate constraints. 
Additionally, the parallel processing capabilities of the GA significantly enhance the efficiency of the approach, making it well-suited for large-scale resource optimization scenarios. 
To further accelerate convergence and avoid the resource optimization problem resulting in local optima, we integrate a trust region method (TRM)-based local optimization component. 


\subsection{KANs Approximation}

In the performance function approximation phase, KANs are used to learn the relationship $P(X)$ between each slice's performance metrics and its control actions. Our goal is to achieve accurate approximation while maintaining the interpretability of the connection between input $X_i$ and system performance output $p_i$. This can be expressed as:
\begin{equation}
    P(X_i) = \text{KANs}(X_i, p_i), \quad i \in I.
\end{equation}
The KANs are founded on the Kolmogorov-Arnold representation theorem \cite{braun2009constructive},\cite{kolmogorov1957representation} as an alternative to traditional MLPs-based solutions to learn the performance function. KANs implement activation functions on edges rather than on nodes, as in MLPs. This unique characteristic enables it to achieve better approximation (e.g., $P(X)$) in terms of accuracy and interpretability.

\textbf{Kolmogorov-Arnold representation theorem.} The theorem~\cite{liu2024kan} asserts that any multivariate continuous function can be expressed as a combination of univariate continuous functions and addition operations. Specifically, for a smooth $f:[0,1]^n \rightarrow \mathbb{R}$,
\begin{equation}
    f(x)=f\left(x_1, \ldots, x_n\right)=\sum_{q=1}^{2 n+1} \Phi_q\left(\sum_{p=1}^n \phi_{q, p}\left(x_p\right)\right),
\end{equation}
where $\phi_{q,p}:[0,1] \rightarrow \mathbb{R}$ and $\Phi_q: \mathbb{R} \rightarrow \mathbb{R}$. Certain univariate functions may be non-smooth or even fractal, rendering them unsuitable for machine learning. KANs extend the aforementioned theorem by generalizing it to arbitrary depths and widths, thereby enhancing their expressive power, which can be represented as follows:
\begin{equation}
\mathrm{KAN}(\mathbf{x})=(\boldsymbol{\Phi}_{L-1} \circ \boldsymbol{\Phi}_{L-2} \circ \cdots \circ \boldsymbol{\Phi}_1 \circ \boldsymbol{\Phi}_0) \circ \mathbf{x}.
\end{equation}
Here, $\boldsymbol{\Phi}_L$ denotes the B-spline function matrix corresponding to the $L$-th KAN layer, and $\mathbf{x}$ is the input vector. The structure of $\boldsymbol{\Phi}$ is given by:
\begin{equation}
\boldsymbol{\Phi}=\left(\begin{array}{ccc}
\phi_{1,1}(\cdot), & \cdots, & \phi_{1, n_{\text {in }}}(\cdot) \\
\vdots & & \vdots \\
\phi_{n_{\text {out }}, 1}(\cdot), & \cdots, & \phi_{n_{\text {out }}, n_{\text {in }}}(\cdot)
\end{array}\right),
\end{equation}
The elements $\phi(x)$  are activation functions, which are defined as the sum of a bias function $b(x)$ and a spline function, weighted by a coefficient $w$:
\begin{equation}
    \phi(x) = w \left(b(x) + \operatorname{spline}(x)\right). 
\end{equation}

KANs are a combination of spline functions and MLPs. This hybrid structure enables KANs to not only learn features but also obtain a high-precision optimization of these features (owing to its internal similarity to spline functions). As a result, the accuracy of our performance function is enhanced. Furthermore, KANs exhibit excellent interpretability. Their inherent visualization and interactivity facilitate human understanding of model behavior and outcomes. This transparency enables better comprehension of the relationships between slice configuration decisions and slice performance during the learning process, providing valuable insights into the underlying system dynamics and decision-making.

\subsection{Problem Optimization}
In the cost minimization phase, the objective is to solve the previously established optimization problem, now mathematically defined by the KAN$-$derived performance functions. This problem involves a large number of variables (e.g., $|R| \times |I|$) and numerous complex constraints (e.g., $|I|+|R|^2+|R|$). Traditional optimization approaches, such as gradient descent (also with unknown convexity) and Bayesian optimization, are often inadequate for solving large-scale optimization problems of this nature. Moreover, some methods are prone to converging to local optima. To overcome these challenges, we employ a genetic method embedded with a trust region approach to navigate the search space and seek a global optimum. 

\textbf{Genetic Method.} The genetic method is inspired by the principles of natural evolution. The evolutionary process maintains diversity among individuals within a population. Individuals that are better adapted to the environment are more likely to survive, reproduce, and pass their traits to the next generation. In the context of evolution, a fitness function is used to evaluate individuals. These individuals undergo selection, crossover, and mutation operators, which simulate the process of biological evolution. In our work, the objective function is mapped to the fitness function $F(\cdot) = C(w_r , x_{i,r}) $. Following the evaluation, the tournament selection \cite{miller1995genetic} is used to choose parents for reproduction and generation of the next population. Next, we apply a crossover operation with a predefined crossover probability. Let $p_1$ and $p_2$ represent two parent individuals, which produce two offspring $c_1$ and $c_2$ through the following crossover formula:
\begin{equation}
\begin{aligned}
& c_1=\alpha \cdot p_1+(1-\alpha) \cdot p_2, \\
& c_2=\alpha \cdot p_2+(1-\alpha) \cdot p_1,
\end{aligned}
\label{eq:crossover}
\end{equation}
where $\alpha$ is a random number, ensuring diversity in the offspring by balancing the traits inherited from both parents. Subsequently, offspring undergo a mutation operation with an adaptive mutation rate $m(g)$ and random number $r$. The mutation is performed as follows:
\begin{equation}
x_i^{\prime}= \begin{cases}x_i+\epsilon & \text { if } r<m(g) \text { and } x_i^{\prime} \text { is feasible, } \\ x_i & \text { otherwise, }\end{cases}
\label{eq:mutation}
\end{equation}
where $x_i^{\prime}$ is the mutated solution after boundary correction and feasibility checks. The $\epsilon$ is the mutation magnitude, following a normal distribution. The adaptive mutation rate is defined as $m(g) = m_0 \cdot \left(1 - {g}/{G}\right)$, where $g$ is the current generation index, $G$ is the total number of generations, and $m_0$ is the initial mutation rate. This strategy ensures a higher mutation rate in the early stages to promote exploration, and a reduced mutation rate in later stages to enhance exploitation and convergence towards the optimal solution.

\textbf{Trust Region Searching.} It defines a localized region around the current iteration $k$. In this region, the objective function $f(x_k)$ is approximated with its second-order Taylor expansion to construct a local quadratic model $m_k(s)$. The TRM then minimizes the quadratic model within the trust region, which can be formulated as follows:
\begin{align}
    &  \min_{s} \quad m_k(s)=f\left(x_k\right)+\nabla f\left(x_k\right)^T s+\frac{1}{2} s^T B_k s \\
    &\text{s.t.} \quad \quad \quad 
     \quad  \|s\| \leq \Delta_k,
     \label{eq:trm}
\end{align}
where $\nabla f(x_k)$ represents the gradient, $B_k$ is the Hessian matrix, and $\Delta_k$ is the radius of the trust region. After each iteration, the algorithm adjusts the region's radius based on the ratio $\rho_k$ between the actual reduction and the predicted reduction in the objective function, defined as:
$\rho_k
 = (f(x_k) - f(x_k + s_k))/(m_k(0) - m_k(s_k))$. If $\rho_k$ approaches 1, it indicates the model closely approximates the objective function, and the current trust region radius is acceptable. Conversely, if $\rho_k$ is close to 0, it suggests a discrepancy between the model and the objective function. In this case, the region radius needs to be reduced in the next iteration.

\begin{algorithm}[!t]
    \caption{The \textit{InSlicing} Algorithm}
    \label{alg:pseudo_code}
    \KwIn{$I, R, Q_i, L_r, H_r, w_r, G,n$}
    \KwOut{$X_{best}, F_{best}$}

    $/**\; KANs \; Approximation \; Process \; **/$\;

    \For{$i = 0, 1, \ldots, |I|$}{
        Initialize KAN models for slice $i$\;
        Train models to learn performance function $P(X_i)$ using online querying\;
    }

    $/**\; Optimization \; Search\; Process \; **/$\;

    Initialize population and parameters for GA\;
    Set best solution $X_{best} \gets \emptyset$, best value $F_{best} \gets \infty$\;

    \For{$g = 0, 1, \ldots, G$}{
        $/**\; Evaluate \; Fitness\; **/$\;
        \ForEach{individual $X$ in population}{
            Calculate fitness function $F(X)$\;
            \If{$F(X) < F_{best}$ and $X$ is feasible}{
                $X_{best} \gets X$, $F_{best} \gets F(X)$\;
            }
        }
        $/**\;Natural \; Evolution\; **/$\;
        Perform selection, crossover and mutation operators with Eq. \ref{eq:crossover} and Eq. \ref{eq:mutation}\;
        
        $/**\; Trust\; Region\; **/$\;
        \If{$g ~\%~ n == 0$ and $X_{best} \neq \emptyset$}{
            Set start point with $X_{best}$\;
            \For{$k = 0, 1, \ldots, K$}{
             $X_{k} \gets$ Solve approximated problem $m_k(s)$\;
             Adjust region radius $\Delta_k$ based on $\rho_{k}$\;
            }
            Update $X_{best} \gets X_{K}, F_{best} \gets F(X_{K})$\;
        }
    }
\end{algorithm}

\subsection{ The \emph{InSlicing} Algorithm}

Based on the preceding analysis, we summarize the \emph{InSlicing} algorithm in Alg. \ref{alg:pseudo_code}. 
In the KANs approximation phase, we use KAN models to learn the mathematical expressions for both control actions and performance metrics for each slice. This transforms the unknown constraints to well-defined formulations. Next, we implement a hybrid optimization framework that combines global exploration with local refinement. we initialize the population with a genetic method, leveraging the cost function as the fitness function to evaluate the individual solutions. Through heuristic search, the GA performs a global exploration to identify a rough optimal solution. Meanwhile, a trust region is defined around the solution. Within this trust region, a quadratic model is constructed, and gradient information is used for local optimization. This dual-phase optimization framework significantly enhances both optimization precision and convergence efficiency.

\section{Evaluation}




In this section, we extensively evaluate the proposed \emph{InSlicing} algorithm in terms of slicing performance, scalability, and regret minimization. 
We use the same end-to-end network testbed from Atlas~\cite{liu2022atlas}, which consists of OpenAirInterface radio access network and core network, and multiple network slices. 
Each slice has one smartphone with a customized Android mobile application, which generally sends different kinds of data to the core and gets feedback accordingly. 
The slice configuration includes \emph{bandwidth\_ul}, \emph{bandwidth\_dl}, \emph{mcs\_offset\_ul}, \emph{mcs\_offset\_dl}, \emph{backhaul\_bw}, and \emph{cpu\_ratio}.
Without loss of generality, the performance metric of all slices is round-trip latency (more details see Atlas~\cite{liu2022atlas}). 
By default, we configure 9 slices with performance thresholds (i.e., $Q_i, \forall i$) [400, 500, 60]ms, and configuration values ranging between [0,1].
The KAN models are learned for 1000 iterations. 
For the optimization process, we set a maximum search time of 400 seconds. 
The genetic method is configured with a population size of 50, a crossover probability of 0.9, and a base mutation probability of 0.2. 
Every 5 evolutions, the best-performing individual is selected as the starting point for trust region optimization, which is configured with a base radius of 0.2 and executes for 25 iterations. 
A clipping operation is applied during the optimization process to ensure that the solutions remain within the resource bounds. 

We compare our proposed algorithm with the following methods:
\begin{itemize}
    \item \textbf{GBO}: The GBO employs a global Bayesian optimization framework (similar to the algorithm in \cite{ayala2023edgebol}) to address the slice configuration problem across all slices. A Gaussian process is used as the surrogate model, and expected improvement is utilized as the acquisition function. To ensure the SLAs of slices, the objective function incorporates penalty terms for performance requirement violations. 
    \item \textbf{GA-only}: The GA-only uses a genetic method based on population evolution principles. It is well-suited for optimization tasks involving a large number of parameters and complex mathematical representations. GA-only focuses solely on global optimization and it cannot perform any local refinement within local regions around promising solutions. 
\end{itemize}

\begin{figure}[!t]
	\centering
	\includegraphics[width=2.7in]{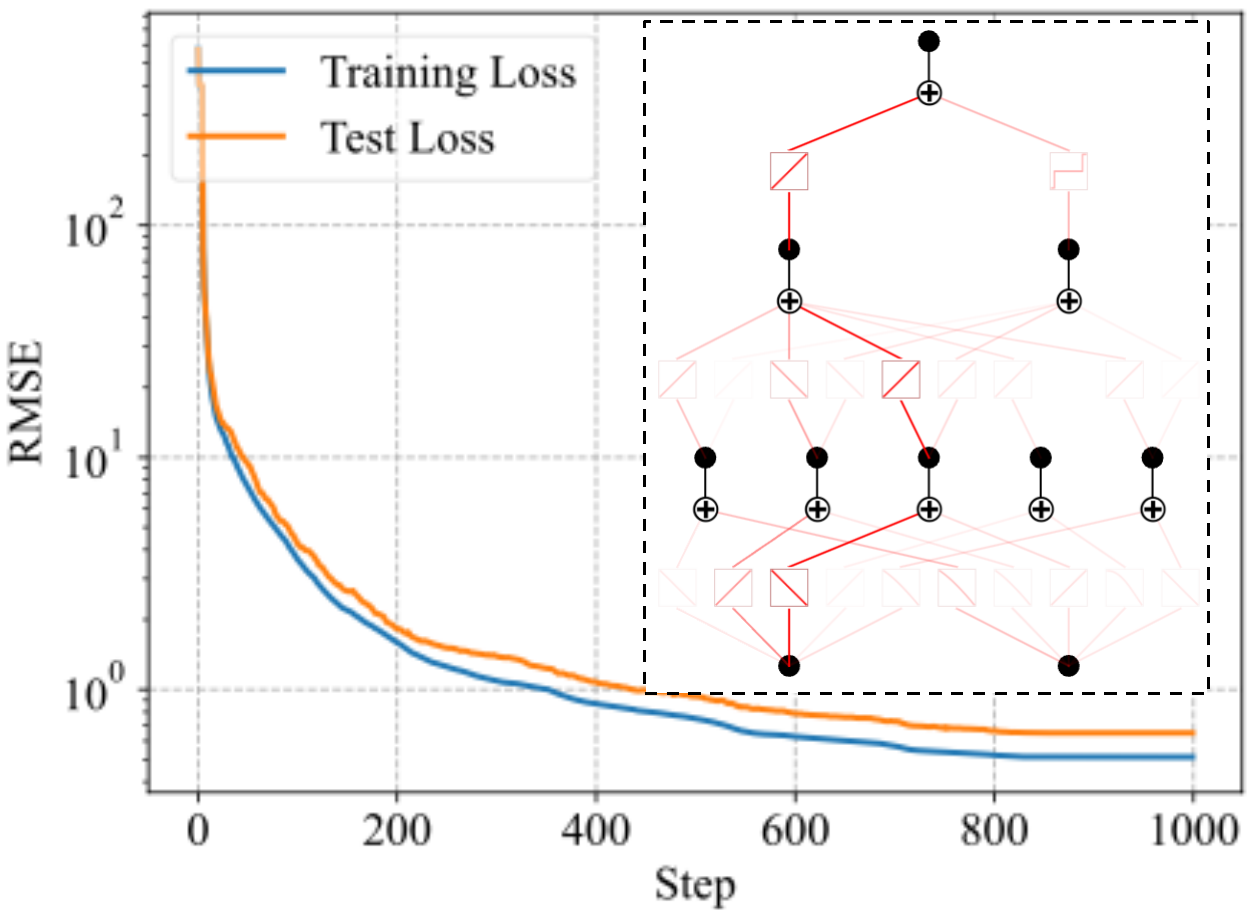}
	\vspace{-0.07in} \caption{\small The learning loss and illustration of KANs.}
	\label{fig:kan_loss_model}
\end{figure}

\begin{figure}[!t]
	\centering
	\includegraphics[width=2.7in]{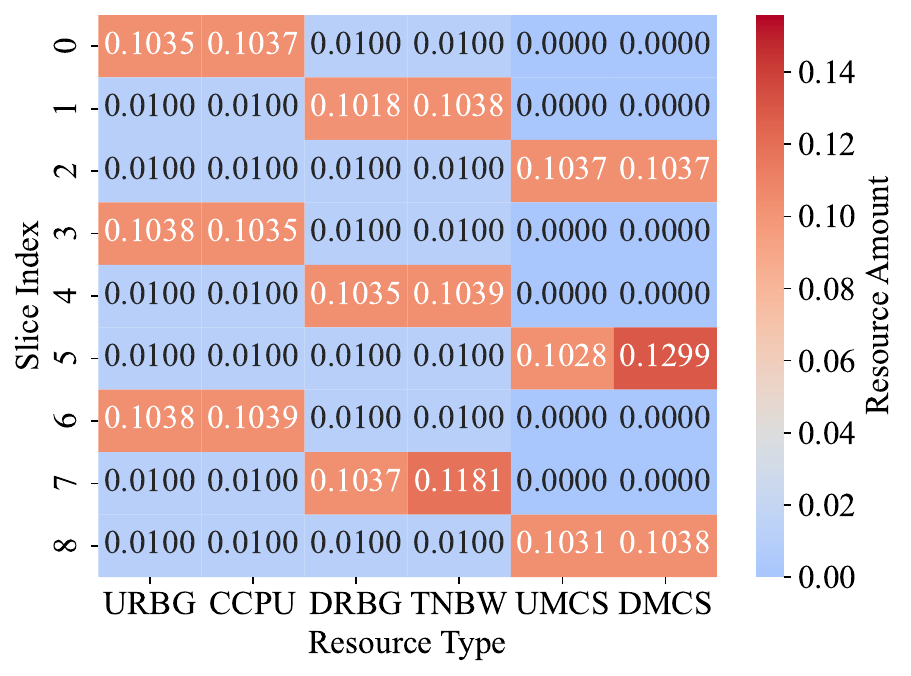}
	\vspace{-0.07in} \caption{\small The heatmap of slice configuration in \textit{InSlicing}.}
	\label{fig:resource_allocation_map}
\end{figure}

\begin{figure}[!t]
	\centering
	\includegraphics[width=3.0in]{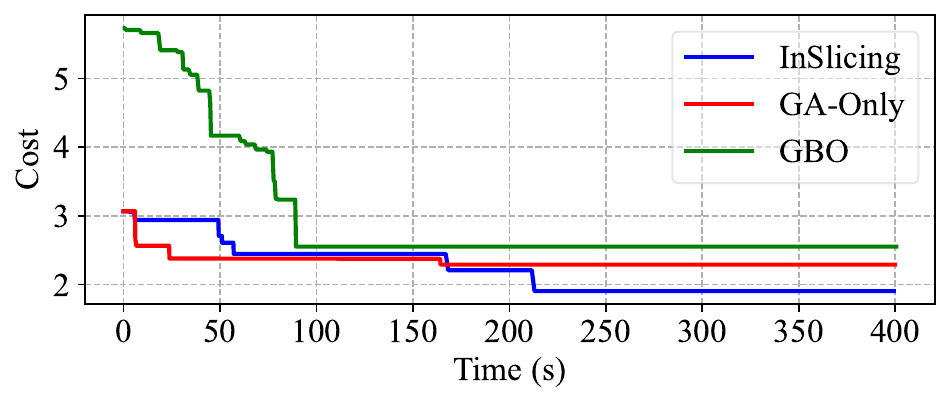}
	\vspace{-0.07in} \caption{\small The cost under comparison algorithms.}
	\label{fig:comparison}
\end{figure}

Fig. \ref{fig:kan_loss_model} illustrates the process of KAN model learning the performance metric for a specific slice. The model begins to converge after approximately 400 steps, achieving low root mean square error (RMSE) values for both training and test losses (0.2867 and 0.3354, respectively). The right side shows the model architecture, which processes two distinct resource inputs through three hidden layers to generate performance predictions. This structure captures the differential impact of the two resources on the performance metric through a series of activation functions. The resulting mathematical expression is formulated as $P(x_1,x_2) = -788.9124x_1 + 11.1672x_2 + 65.9526 \sin(0.6438x_1 - 4.1988) - 154.8258 \sin(0.5912x_2 + 5.1804) + 45.7148 \sin(0.6943x_2 + 5.2315) + 15.8861 \sin(1.9874x_2 + 7.6146) + 836.0928$.

Fig. \ref{fig:resource_allocation_map} illustrates the optimization results of the proposed \textit{InSlicing} to minimize the cost while ensuring that slice users' performance requirements are satisfied. The figure depicts the slice configuration map across 9 slices for 6 resource types. To maintain the operation of  MAR and HVS applications, the minimum resource thresholds for URBG, CCPU, DRBG, and TNBW are set to 0.1.

Fig. \ref{fig:comparison} compares the proposed \textit{InSlicing} with other methods, demonstrating that our approach achieves the lowest MNOs operational cost of 1.90 under the same computational time complexity, outperforming both GBO (2.55) and GA-only (2.28) approaches. This is because GBO struggles with high-dimensional spaces due to the sparsity of sampled points. Moreover, surrogate models like Gaussian processes often fail to accurately approximate complex functions in high-dimensional spaces, particularly those with oscillations or multiple local optima. This leads to insufficient guidance for the optimization search (e.g., the cost is more than twice that of the other algorithm at the beginning of the optimization process), causing it to converge prematurely to local optima. The GA-only utilizes population-based search strategies, which makes it well-suited for problems with multiple local optima and high-dimensional spaces, resulting in better solutions than GBO. However, as the population evolves, diversity can decrease, and individuals also get trapped in local optima. Our proposed \textit{InSlicing} incorporates local refinement using the trust region. While GA provides a rough global solution, TRM leverages gradient information within the trust region to further optimize the solution locally, where the hybrid approach mitigates the degradation of global search capability caused by reduced population diversity.

\begin{figure}[!t]
	\centering
	\includegraphics[width=3.0in]{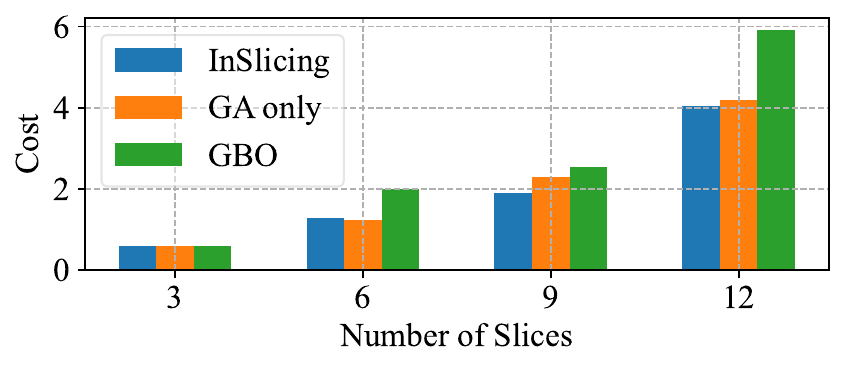}
	\vspace{-0.07in} \caption{\small The scalability under comparison algorithms.}
	\label{fig:scalability}
\end{figure}

\begin{figure}[!t]
	\centering
	\includegraphics[width=3.0in]{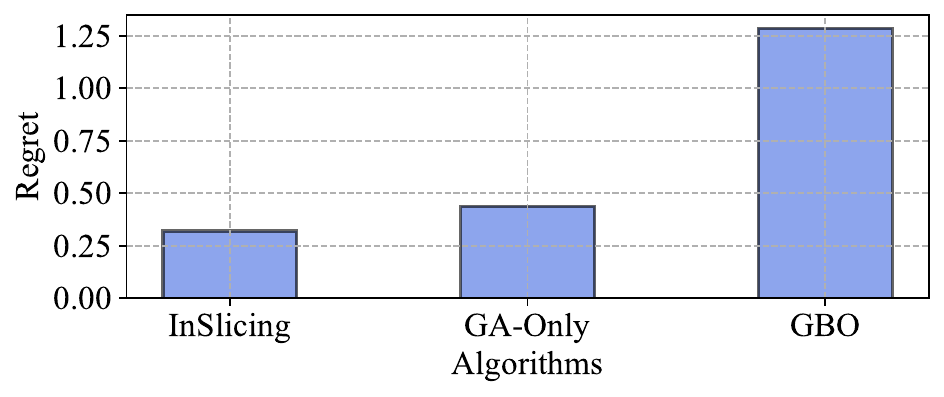}
	\vspace{-0.07in} \caption{\small The regret under comparison algorithms.}
	\label{fig:regret}
\end{figure}

\begin{figure}[!t]
	\centering
	\includegraphics[width=3.0in]{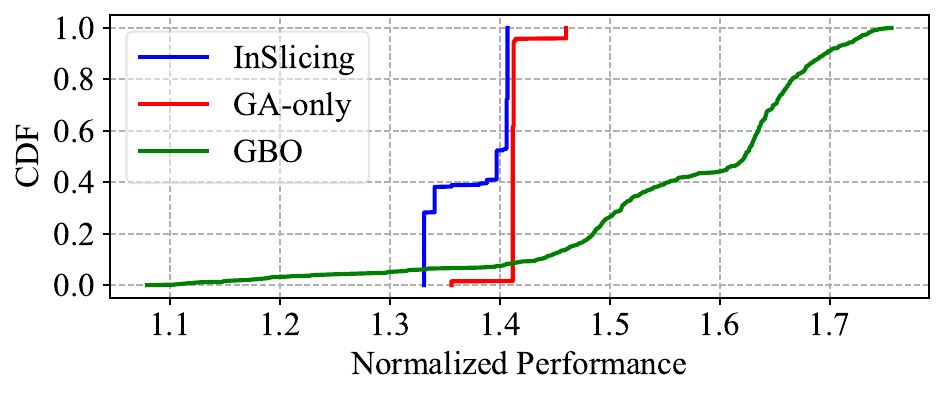}
	\vspace{-0.07in} \caption{\small The CDF of normalized performance.}
	\label{fig:performance_cdf}
\end{figure}

Fig. \ref{fig:scalability} compares the scalability of various algorithms under different numbers of slices in the network. It shows all three algorithms are capable of handling resource orchestration under different slice configurations. As the number of slices increases, the optimal cost also rises correspondingly. Our proposed \textit{InSlicing} can achieve the lowest cost among the three methods (at six slices, although our optimization result is not the absolute lowest, it remains highly competitive and approaches the minimum). Interestingly, GBO always obtains higher costs, likely due to its difficulty in handling high-dimensional optimization problems. 

Fig. \ref{fig:regret} compares the regret values under the default settings. The regret value is defined as the ratio between the cumulative gap (the sum of differences between the cost value after each optimization iteration and the final optimal cost value) and the number of iterations. The results show that our proposed \textit{InSlicing} achieves the lowest regret value, indicating its ability to significantly reduce the gap between the objective function value and the optimal value throughout most iterations. In contrast, GBO exhibits the highest regret value, which can be attributed to its premature convergence to a suboptimal solution as shown in Fig. \ref{fig:comparison}.

Fig. \ref{fig:performance_cdf} presents the cumulative distribution function (CDF) of normalized performance under the three algorithms. It can be observed that in most cases (e.g., normalized performance ranges from 1.33 to 1.41), our algorithm obtains lower normalized performance, which indicates that our algorithm precisely meets slice users’ requirements avoiding unnecessary resource overprovisioning. In comparison, the curve of GBO shows a more dispersed distribution. This suggests that GBO tends to over-allocate resources to meet slice users' requirements and fails to effectively control the MNOs' costs in Fig. \ref{fig:comparison}.

\section{Related work}

\textbf{Resource Management in Network Slicing.} A primary challenge in network slicing is the efficient allocation of resources to support heterogeneous user services. Leconte et. al. \cite{leconte2018resource} proposed a framework for fine-grained slice configuration in terms of network bandwidth and cloud processing. Their approach considers traffic fairness and computational fairness, utilizing an alternating direction method of multipliers (ADMM)-based iterative algorithm, which is proven to converge to optimal resource allocation. Fossati et. al. \cite{fossati2020multi} investigated the fair sharing among slices under insufficient resources scenarios. They proposed an optimization framework based on the ordered weighted average (OWA) operator to ensure fairness in both single-resource and multi-resource allocation problems. The O-RAN \cite{polese2023understanding} advocates for an open network structure and promotes the use of open RAN interfaces to interconnect components with a RAN intelligent controller (RIC) for managing and controlling resources. The OnSlicing \cite{liu2021onslicing} proposed an online end-to-end network slicing system based on the O-RAN. This system aimed to minimize resource usage while avoiding violations of slice SLAs and infrastructure resource capacity. By leveraging domain managers, it achieved subsecond-level control for resource allocation. However, existing work often overlooks the cost of resources. In contrast, our work approaches the problem from the perspective of MNOs, focusing on minimizing operation costs while ensuring SLA compliance for slice users. 

\textbf{AI/ML for Networking.} AI/ML techniques have emerged as powerful tools for network orchestration due to the complex nature of network environments and intricate interactions between multiple network components. It provides adaptive and data-driven solutions that can accommodate these variations. DeepSlicing \cite{liu2020deepslicing} proposed an efficient resource allocation framework that decomposes the optimization problem into multiple sub-problems. Their approaches combine convex optimization for the master problem and deep deterministic policy gradient (DDPG) agents for slave problems, aiming to maximize utility functions. Li et. al. \cite{li2018deep} explored two typical demand-aware slice configuration scenarios, wireless resource slicing and priority-based core network slicing. Compared to demand-prediction-based approaches, their DRL framework could implicitly capture deeper relationships between demand and supply, improving the effectiveness and flexibility of network slicing. Atlas \cite{liu2022atlas} proposed an online network slicing system that leverages Bayesian optimization to reduce the sim-to-real discrepancy. Using Gaussian process regression, the system learns real-world policies online, enhancing the performance of network slicing strategies. 
However, existing solutions are mostly based on DNN-based model representations, which lack transparency and interpretability, and thus constrain their practical deployment in real-world networks.

\section{Conclusion}
In this paper, we presented a new interpretable network slice configuration algorithm in open radio access networks, by integrating Kolmogorov-Arnold Networks (KANs) and hybrid optimization process.
On the one hand, we use KANs to approximate and learn the unknown performance
function of individual slices, which converts the blackbox optimization problem. On the other hand, the genetic method embedded with trust region is used to solve the converted problem.
With the extensive evaluation, we show that our proposed algorithm achieves high interpretability while reducing 25+\% operation cost than existing solutions.
We believe this work provides a new perspective of slice configuration, by considering the solution interpretability beyond the solution performance only.


\bibliographystyle{IEEEtran}
\bibliography{ref/reference}

\end{document}